\documentclass[
reprint,
superscriptaddress,
 amsmath,amssymb,
prl,
floatfix,
]{revtex4-1}

\usepackage{float}
\usepackage{graphicx}
\usepackage{xcolor}
\usepackage{amsmath,amssymb,amsthm,amsfonts}
\usepackage{bbm}
\usepackage{bm}
\renewcommand{\vec }{\bm}
\renewcommand{\mathbf}{\bm}
\usepackage{enumerate}

\usepackage{dsfont}
\begin{document}
\title{
Topological surfaces
of domain wall-decorated antiferromagnetic \\ topological insulator
MnBi$_{2n}$Te$_{3n+1}$
}

\author{Yihao Lin}
\affiliation{International Center for Quantum Materials, School of Physics, Peking University, Beijing 100871, China}

\author{Ji Feng}
\email{jfeng11@pku.edu.cn}
\affiliation{International Center for Quantum Materials, School of Physics, Peking
University, Beijing 100871, China}
\begin{abstract}
Antiferromagnetic topological insulators harbor topological in-gap surface states protected by an anti-unitary $S$ symmetry, which is broken by the inevitable presence of domain walls. Whether an antiferromagnetic topological insulator with domain walls is gapless and metallic on its topological surfaces remains to be elucidated. We show that a single non-statistical index characterizing the magnetic order of domain wall-decorated antiferromagnetic topological insulator, referred to as the Ising moment, determines the topological surface gap, which can be zero even when the $S$ symmetry is manifestly broken. In the thermodynamic limit, the topological surface states tend to be gapless when magnetic fluctuation is bounded. In this case, the Lyapunov exponent of the surface transfer matrix reveals a surface delocalization transition near the zero energy due to a crossover from orthogonal to chiral orthogonal symmetry class. Spectroscopic and transport measurements on the surface states will reveal the critical behavior of the transition, which in return bears on the nature of antiferromagnetic domains walls.

\end{abstract}

\pacs{}
\maketitle
\textcolor{blue}{\textit{Introduction.}}
 As a representative example of magnetic topological crystalline insulator,\cite{zhang2015}
the antiferromagnetic (AFM) topological insulator (TI) is characterized by a $\mathbb{Z}_2$ invariant protected by a composite anti-unitary symmetry $S=\Theta T_{1/2}$, where $\Theta$ is time-reversal and $T_{1/2}$ is a half lattice translation. Topologically protected gapless states are expected on surfaces respecting the $S$ symmetry. Recently, a family of layered compounds with an intrinsic A-type interlayer AFM order, $\mathrm{MnBi_{2n}Te_{3n+1}}$, have been shown to be AFM TIs\cite{otrokov2019}, whose thin films can display quantum anomalous Hall and topological magnetoelectric effects.~\cite{li2019,zhang2019b,liu2020a,deng2020a,ge2020} The surface spectrum measurements have focused on the top surface\citep{otrokov2019,vidal2019a, hao2019,swatek2020a} but the topological surface states on symmetry-preserving surfaces have so far evaded detection. Lattice imperfections, such as surfaces and defects, can violate the $S$ symmetry and lead to bandgap or localization of the surface states.\cite{ando2015,zhang2019b,li2019,deng2020a} Of particular interest for a layered AFM TI is the inevitable presence of antiferromagnetic domain walls comprised of a ferromagnetic bilayer, which generally breaks the $S$ symmetry. Then whether the topological surface states remain gapless and/or conductive in an AFM TI with symmetry-breaking domain walls requires elucidation. 


\begin{figure}[b]
\includegraphics[scale=0.9]{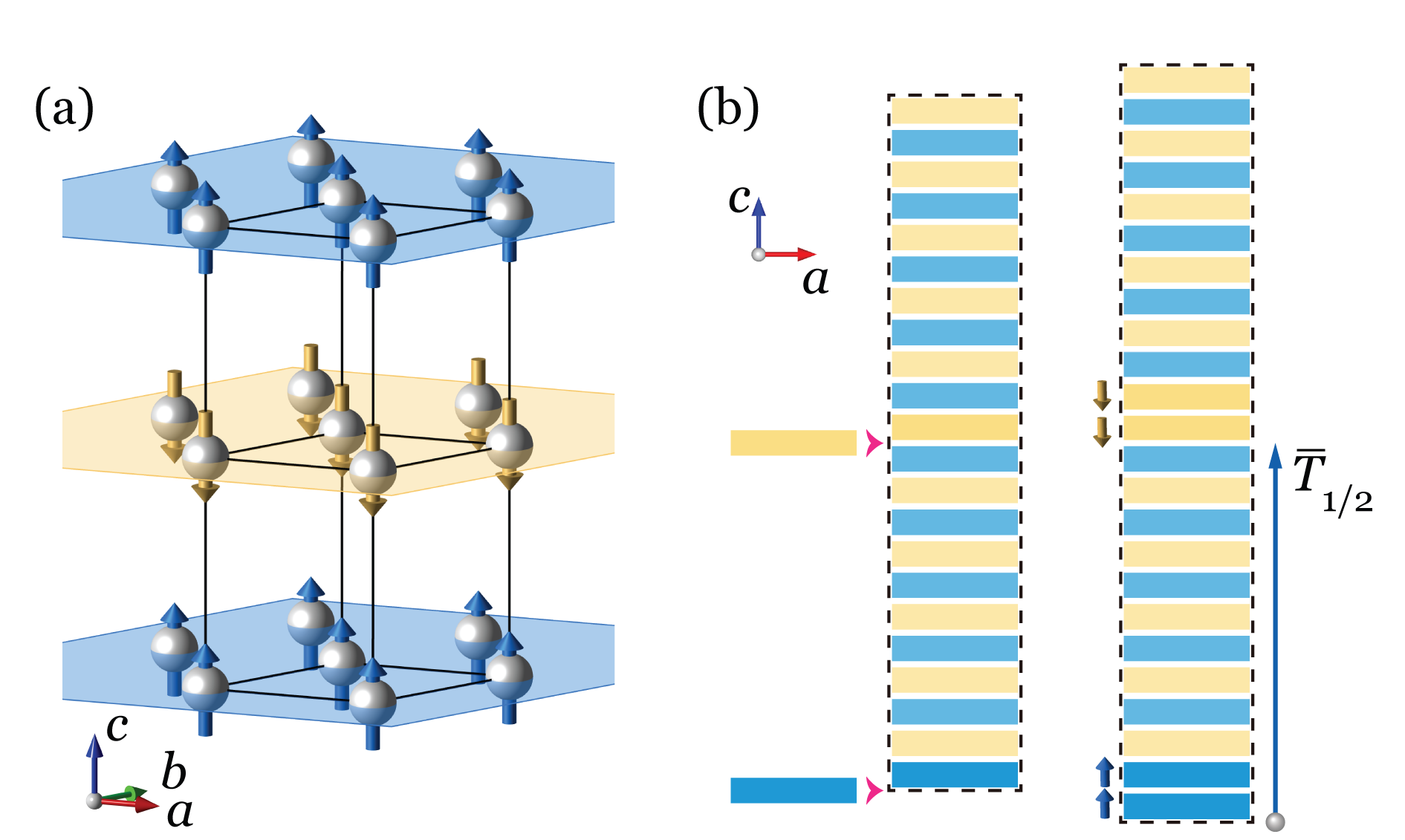}
\caption{The effective tight-binding model of layered antiferromagnetic MnBi$_{2n}$Te$_{3n+1}$. (a) One magnetic unit cell is comprised of two layers of hexagonally packed sheets of Mn$^{2+}$ with opposite magnetization. $\mathbf{a}$ is along x-axis. $\mathbf{c}$ is along z-axis.  (b) A ferromagnetic bilayer as a domain wall.}
\label{fig:01}
\end{figure}

In this paper, we investigate the surface spectrum and transport properties based on an electronic model for a domain wall-decorated AFM TI and found the presence of domain walls has an intriguing impact on the topological surface states. 
Remarkably,	the topological surface gap is a function of a single non-statistical variable describing the overall magnetic order, which is termed the \textit{Ising moment}. The surface spectrum can be gapless even when the requisite $S$ symmetry is broken by domain walls. In the thermodynamic limit, the Ising moment converges to special gapless values when magnetic fluctuation is bounded. Computed Lyapunov exponents of surface transfer matrices show that the topological surface states are generally localized, except with an emergent chiral symmetry a delocalization transition occurs at a single energy. With bounded magnetic fluctuation, the crossover to chiral symmetry is broadened and displays an unconventional critical scaling, providing experimentally accessible transport signatures that in return uncover the nature of antiferromagnetic domains walls.

\textcolor{blue}{\textit{Surface gap and Ising moment}}. 
We will suppose that our model system is topologically nontrivial according to the $S$ symmetry in a perfect AFM configuration, as shown in Fig.\ref{fig:01}(a).
Although the presence of domain walls generally breaks the $S$ symmetry and nullifies the $S$-protected topology, in a supercell a composite symmetry $\bar{S}=\Theta\bar{T}_{1/2}$ may emerge. This is the case if domain walls of opposite magnetization alternate with equal spacings, which can be achieved by inserting magnetic monolayers into a perfect AFM supercell as shown in Fig.\ref{fig:01}(b). If this insertion process can be seen as a continuous deformation preserving $\bar S$ symmetry, the resultant domain wall superlattice is bound to have gapless surfaces. The more relevant and interesting question is what happens to the surface spectrum when domain walls are randomly placed with random magnetizations.



\begin{figure}[h]
    \centering
    \includegraphics[width=8cm]{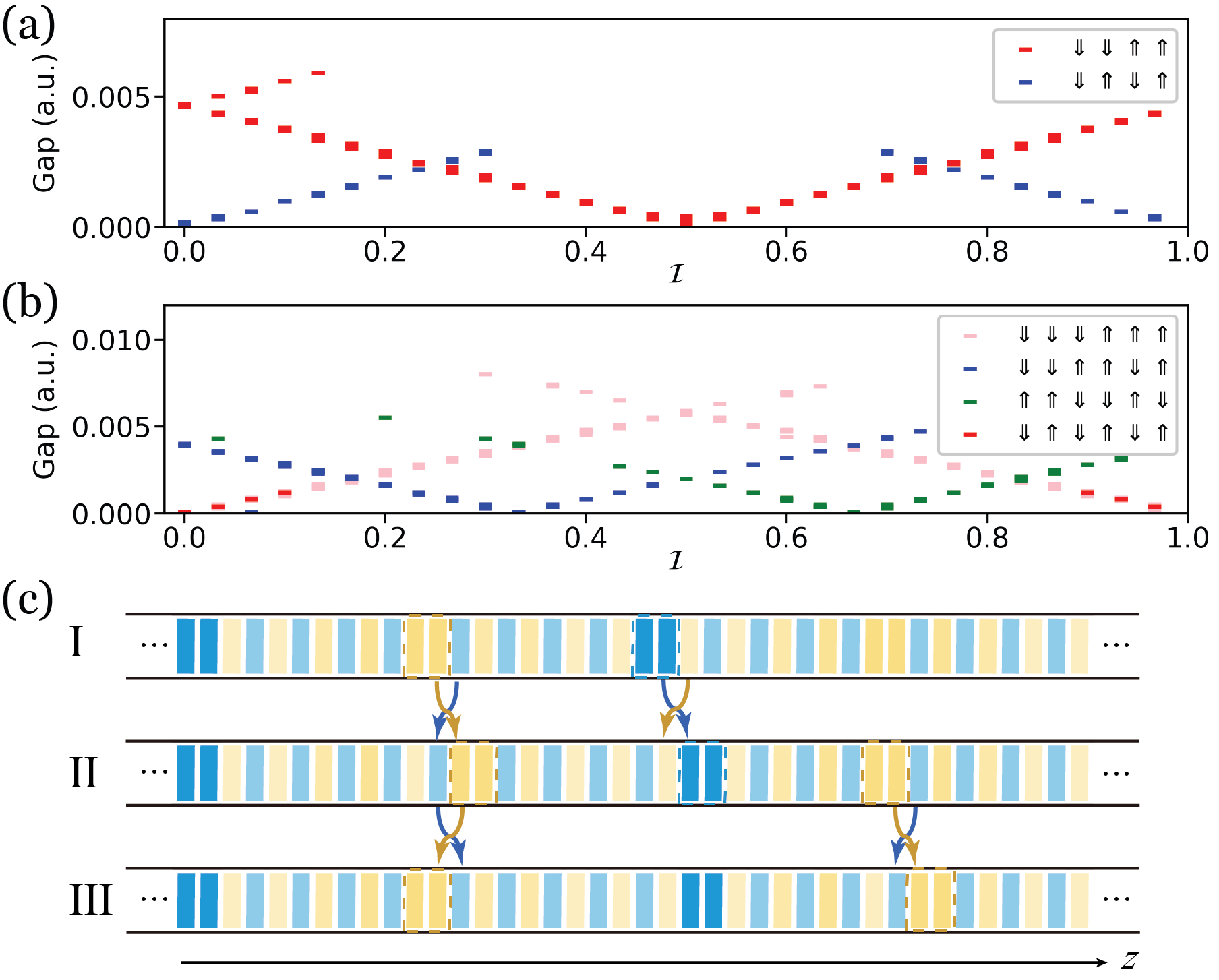}
    \caption{Surface gaps of 60-layer supercells with (a) 4 domain walls and (b) 6 domain walls, plotted against the Ising moment $\mathcal{I}$. (c) Migrations of domain walls that will keep $\mathcal{I}$ invariant, namely, I$\rightarrow$II and II$\rightarrow$III.}
    \label{fig:03}
\end{figure}

An effective tight-binding model is available for $\mathrm{MnBi_{2n}Te_{3n+1}}$, from regularizing the 4-band $k\cdot p$ model of the  topological insulator Bi$_2$Se$_3$\cite{zhang2009} on a hexagonal lattice (Fig. \ref{fig:01}(a)) and adding a layer-staggered Ising-type exchange field describing the AFM order.\cite{zhang2020} We generalize this model to include a layer-dependent Ising field in the Hamiltonian
\begin{equation}
    H=\sum_{\ell\vec q} \psi_{\ell\vec q}^\dagger D_{\vec q}(m_\ell)\psi_{\ell\vec q} +\psi_{\ell\vec q}^\dagger J_{\vec q}\psi_{\ell+1\vec q} +\psi_{\ell+1\vec q}^\dagger J_{\vec q}^\dagger\psi_{\ell\vec q},
    \label{eq:hamil}
\end{equation}
where integer $\ell$ is the magnetic  monolayer index and $\vec q=(k_x,k_y)$ is the in-plane wavevector. Bloch bases in each magnetic monolayer are used, and the field operator $\psi_\ell$ and matrices $J_{\vec q}$ and $D_{\vec q}(m_l)$ have dimensions of 4. The intralayer exchange interaction is modulated by the Ising order parameter $m_{\ell} =\pm 1$. For instance, $m_\ell= (-1)^\ell$ for the perfect AFM ordering. Domain walls can be easily introduced by $m_\ell$. Descriptions of the parameters, electronic structure and symmetry of this model are provided in the Supplemental Materials (SM).\cite{suppl} It suffices to mention here that the parameters are chosen to ensure the system is topological in presence of $\bar S$ symmetry.\cite{suppl} 
For an unbiased survey of the effect of domain wall decoration on surface states,  a series of supercells made of $N=60$ layers with randomly placed $n$ domain walls are examined. Each configuration has a zero net magnetization, which requires an equal number of $\Uparrow$ and $\Downarrow$ domain walls ($n_\Uparrow = n_\Downarrow$). The bandgaps of (010) surface (perpendicular to $y$-axis) states at the Dirac point ($\Gamma$) are calculated using the iterative Green's function method\cite{sancho1985}. Remarkably, the computed  surface gap is seen to be a function of a single variable, which we call the \textit{Ising moment}
\begin{equation}
    \mathcal{I} = \overline{\ell m_\ell}\text{ mod }1,
\end{equation}
where the overbar stands for averaging over all magnetic layers ($\mathcal{I}$ is invariant if calculated in a larger periodic or origin-shifted supercell). This is exemplified in Figs.\ref{fig:03}(a) and (b) with the numbers of domain walls per supercell $n=4$ and $6$, respectively. The surface states turn out to be gapless at special $\mathcal{I}$ values
\begin{equation}
    \mathcal{I}^* = p/n, p=0,2,\cdots, n-2.
    \label{eq:mu*}
\end{equation}
And the specific value of $\mathcal{I}^*$ depends on the permutation of domain wall magnetization in the supercell.

In particular, a configuration without the $\bar S$ symmetry can also be accompanied by gapless (010) surface states. Indeed, each $\mathcal I^*$ corresponds to multiple domain wall configurations modulo cyclic transformations. This can be understood from how $\mathcal{I}$ changes with domain wall migrations. An elementary domain wall migration is accomplished by a local transposition of a pair of magnetic layers, which can be either (A) $\uparrow\downarrow \,\mapsto \,\downarrow\uparrow$ or (B) $\downarrow\uparrow\,\mapsto\, \uparrow\downarrow$. (A)/(B) changes $\mathcal{I}$ by $\pm 2/N$ and thus $\mathcal{I}$ is unaltered when performing transpositions (A) and (B)  on two separate domain walls. As shown in Fig.\ref{fig:03}(c), this makes either a pair of $\Uparrow$ and $\Downarrow$ travel in the same direction (I$\rightarrow$II), or a pair of like domain walls travel in opposite directions (II$\rightarrow$III); in both cases $\mathcal I$ is unchanged.

An pair of $\Uparrow$ and $\Downarrow$ separated \textit{only} by AFM layers can be brought next to each other by a sequence of local transpositions, and then annihilated by an additional transposition. And all domain walls can be removed by consecutive pair annihilations, ending in the perfect AFM order. Keeping track of the changes in $\mathcal{I}$ in this process of approaching the perfect AFM order ($\mathcal{I}=0.5$) provides us with a formula for $\mathcal{I}$, up to modulo 1,
\begin{equation}
    \mathcal{I} = 0.5 + \frac{1}{N} \sum^{n}_{i=1} \alpha_i p_i = \mathcal{I}^* + \frac{1}{N} \sum^{n}_{i=1} \alpha_i \delta p_i  
    \label{eq:mu}
\end{equation}
where $p_i-2$ is the number of magnetic layers between the $i$th and $(i+1)$th domain walls, and $\alpha_i$ is $n_{\Uparrow}-n_{\Downarrow}$ in from the $(i+1)$th to the $n$th domain walls. The second equality in Eq.(\ref{eq:mu}) is derived from the observation that, given a domain wall permutation,  $\mathcal{I}^*$ is attained at equal separations (if possible) i.e. $\delta p_i \equiv p_i-\frac{1}{n}\sum_i p_i=0$.\cite{suppl} This result has interesting consequences in the thermodynamic limit, to be returned to shortly.

\begin{figure}[t]
    \centering
    \includegraphics[width=8cm]{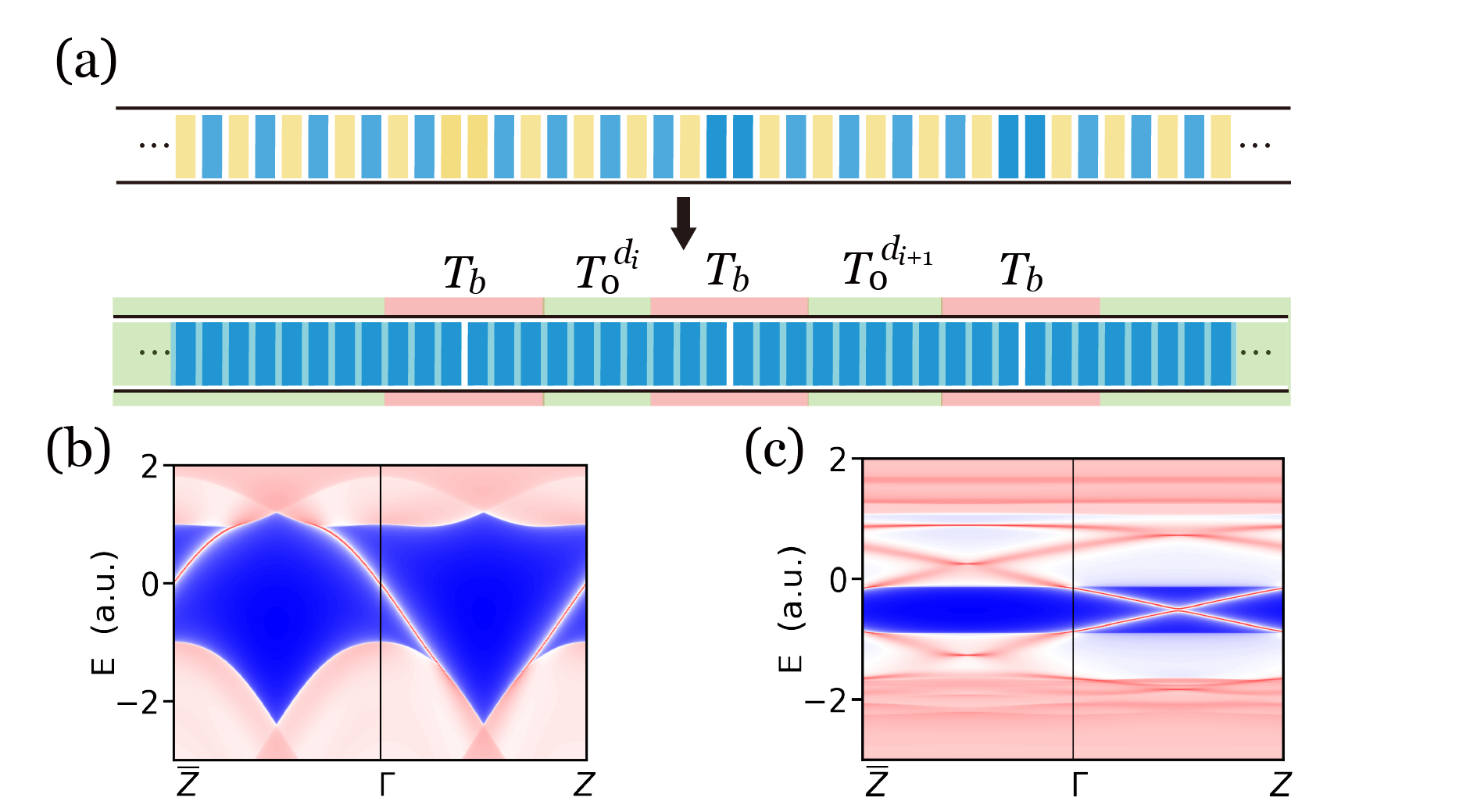}
    \caption{(a) A local gauge transformation changes the AFM domain wall model (top) into the FM bond defect model (bottom). The bond defect zones are highlighted as pink blocks, and FM regions green blocks. (010) surface spectra of (b) the FM bulk without bond defects and (c) a superlattice of defect zones.}
    \label{fig:04}
  \end{figure}

\textcolor{blue}{\textit{Surface transfer matrix}}. We now derive the relation between the surface gap and the Ising moment, with the help of a bond defect model described by the Hamiltonian
\begin{equation}
    H'=\sum_{\ell\vec q} \psi_{\ell\vec q}^\dagger D_{\vec q}(+1)\psi_{\ell\vec q} +\psi_{\ell\vec q}^\dagger J_{\ell\vec q}\psi_{\ell+1\vec q} +\psi_{\ell+1\vec q}^\dagger J_{\ell\vec q}^\dagger\psi_{\ell\vec q},\label{eq:bonddefect}
\end{equation}
where $D_{\vec q}$ is independent of $\ell$, signifying a ferromagnetic order. The interlayer hopping $J_{\ell\vec q}$ depends on the magnetizations of the two layers it connects in the original domain wall model
\[\uparrow\uparrow \text{ or  } \downarrow\downarrow: J_{\ell\vec q}= J_{\vec q}; \; \;\; \uparrow\downarrow \text{ or } \downarrow\uparrow: J_{\ell\vec q}= \mathrm i M_xJ_{\vec q},\]
where $M_x$ stands for reflection about a mirror perpendicular to $x$-axis.
In this model, $J_{\ell \vec q}$ is uniform  ($ \mathrm i M_xJ_{\vec q}$) except bond defects ($J_{\vec q}$) at the original domain walls, as schematized in Fig.\ref{fig:04}(a). It can be verified through symmetry analysis that
the bond defect model Eq.(\ref{eq:bonddefect}) at $k_x=0$ can be obtained from the domain wall model Eq.(\ref{eq:hamil}) by applying $\mathrm i M_x$ to every $\downarrow$ layer, and hence the two models are equivalent at $k_x=0$ apart from a local gauge transformation.\citep{suppl} This bond defect model allows us to analyze the layer-wise transfer matrices of the (010) surface at $k_x=0$, and since the surface Dirac point occurs also at $k_x=0$, the surface transfer matrices devised can be used to analyze the existence of surface gap.

For  Eq.(\ref{eq:bonddefect}), the FM bulk without bond defects has a pair of gapless (010) surface modes, for it corresponds to the perfect AFM state before the gauge transformation. The surface Dirac cone is unfolded in the FM Brillouin zone, as shown in Fig.\ref{fig:04}(b). Thus, the in-gap surface spectra describe a two-channel ballistic conductor with the transfer matrix at an energy
\begin{equation}
    T_0(k_z)=\operatorname{diag}[e^{\mathrm i k_z}, e^{\mathrm i(\pi-k_z)}] =\sigma_zT(k_z),
\end{equation}    
where 
$T(\xi) =\mathrm{diag} [
        e^{\mathrm i\xi} , e^{-\mathrm i\xi}]$,
and $k_z$ is the wavevector of the surface mode moving in $-z$ direction. 

 We define a defect zone comprised of $2L$ layers centered at a bond defect, depicted in Fig.\ref{fig:04}(a).  $L$ is large enough so that the evanescent wave escaping the defect zone is negligible. A superlattice of defect zones supports gapless surface states since it corresponds to an AFM configuration with $\bar S$ symmetry before the gauge transformation. Its surface spectra are shown in Fig.\ref{fig:04}(c), from which we can write down the surface transfer matrix of a defect zone
\begin{equation}\label{Tbond}
    T_b(\theta, u)=
    W(u)T_0(\theta )W(u)^{-1}
\end{equation}
where $\theta$ is the wavevector in the supercell Brillouin zone of the mode propagating in $-z$ direction at a given energy. The matrix $W(v)=[[1,u]^{T},[u^*,1]^{T}]$ accounts for the scattering of the FM surface modes as an electron enters the bond defect. \cite{suppl}

Consider a supercell of $N$ (even) layers with $n$ (even) bond defects. Between the $i$th and $(i+1)$th defect zones there are $d_i$ ferromagnetic layers. The surface transfer matrix of the supercell is then
\begin{eqnarray}
    T_n&=&T_0^{d_n}(k_z) T_b(\theta,u) \cdots  T_0^{d_1}(k_z)T_b(\theta,u) \nonumber \\
    &=&
    \sigma_z^{d_n}T(\delta \phi_n) T_b(\theta',u')\cdots  \sigma_z^{d_1}T(\delta \phi_1) T_b(\theta',u').
    \label{eq:Tn}
\end{eqnarray}
where $\delta \phi_i$ is difference of $\phi_i\equiv d_ik_z$ with its average $\bar{\phi}=\bar{d}k_z$, and in the second line we have introduced   $T_b(\theta',u')= T(\bar \phi)T_b(\theta,u)$.
If the supercell indeed corresponds to an AFM supercell with zero magnetization, then for the energy at which $\theta'=0$, the trace of $T_n$ to second order is found to be~\cite{suppl}
\begin{equation}
    \operatorname{tr}T_n=2 + 16N^2k_z^2|u'|^2 (\mathcal{I}-\mathcal{I}^*)^2,
    \label{eq:traceTn}
\end{equation}
where $\mathcal{I}-\mathcal{I}^*$ is given by Eq.(\ref{eq:mu}) using $\delta \phi_i=k_z\delta p_i$.

As revealed by Eq. (\ref{eq:traceTn}) the trace invariant of the surface transfer matrix, like the surface gap, is a function of $\mathcal{I}$ . This actually is not a coincidence,
since $\det T_n=1$, $T_n$ supports propagating modes when $|\operatorname{tr}T_n|<2$. 
Eq.(\ref{eq:traceTn}) thus indicates a surface gap at $\mathcal{I}\neq \mathcal{I}^*$ 
when $\operatorname{tr}T_n>2$. And at $\mathcal{I}=\mathcal{I}^*$, $\operatorname{tr}T_n=2$ implies that the eigenvalues of $T_n$,
$\lambda_1=\lambda_2=1$, corresponding to a surface Dirac point. Consequently, Eq.(\ref{eq:traceTn}) confirms the empirical relation in Eq. (\ref{eq:mu}), regarding the existence and value of $\mathcal{I}^*$ when a domain wall-decorated AFM TI possesses gapless (010) surface states.
Moreover, the Dirac point of $\mathcal{I}=\mathcal{I}^*$ supercells occurs at $\theta'=0$ implies that the Dirac point energy depends only on the number density of domain walls, and not on their arrangement. 
%

\textcolor{blue}{\textit{Delocalization in thermodynamic limit}}. In the thermodynamic limit where $n\rightarrow\infty$ and $N/n\rightarrow$ const., the localization of (010) surface at $k_x=0$ is characterized by the Lyapunov exponent~\citep{Goldsheid89,Kramer93} of $T_n$:
\begin{equation}
    \gamma = \lim_{n\rightarrow\infty} \frac{\log ||T_n||}{n}
    \label{eq:gamma}
 \end{equation}
which is related to the dimensionless conductance through $g\sim \operatorname{sech} ^2n\gamma$.\citep{Pichard86} We calculate the Lyapunov exponent \cite{Geist90} according to Eq. (\ref{eq:Tn}) with $T^{d_i}_0(k_z)$ calculated as $\sigma^{d_i}_z T(\phi_i)$. On the premise that the domain walls are placed randomly owing to weak interactions, the nearest-neighbor domain wall separations independently follow an identical exponential distribution. Accordingly, $\phi_i$ is sampled as a continuous variable via $\phi_i= \bar \phi x$ with $x$ drawn according to the probability density $p(x)=\exp(-x), x\in[0, \infty)$. 
As only whether $d_i$ is even or odd enters into $\sigma_z^{d_i}$, they are sampled as a binary sequence, fulfilling the condition $n_\Uparrow = n_\Downarrow$. Concerning the domain wall magnetizations, two ensembles (I and II, to be described) have been examined.

\begin{figure}[ht]
    \centering
    \includegraphics[width=8cm]{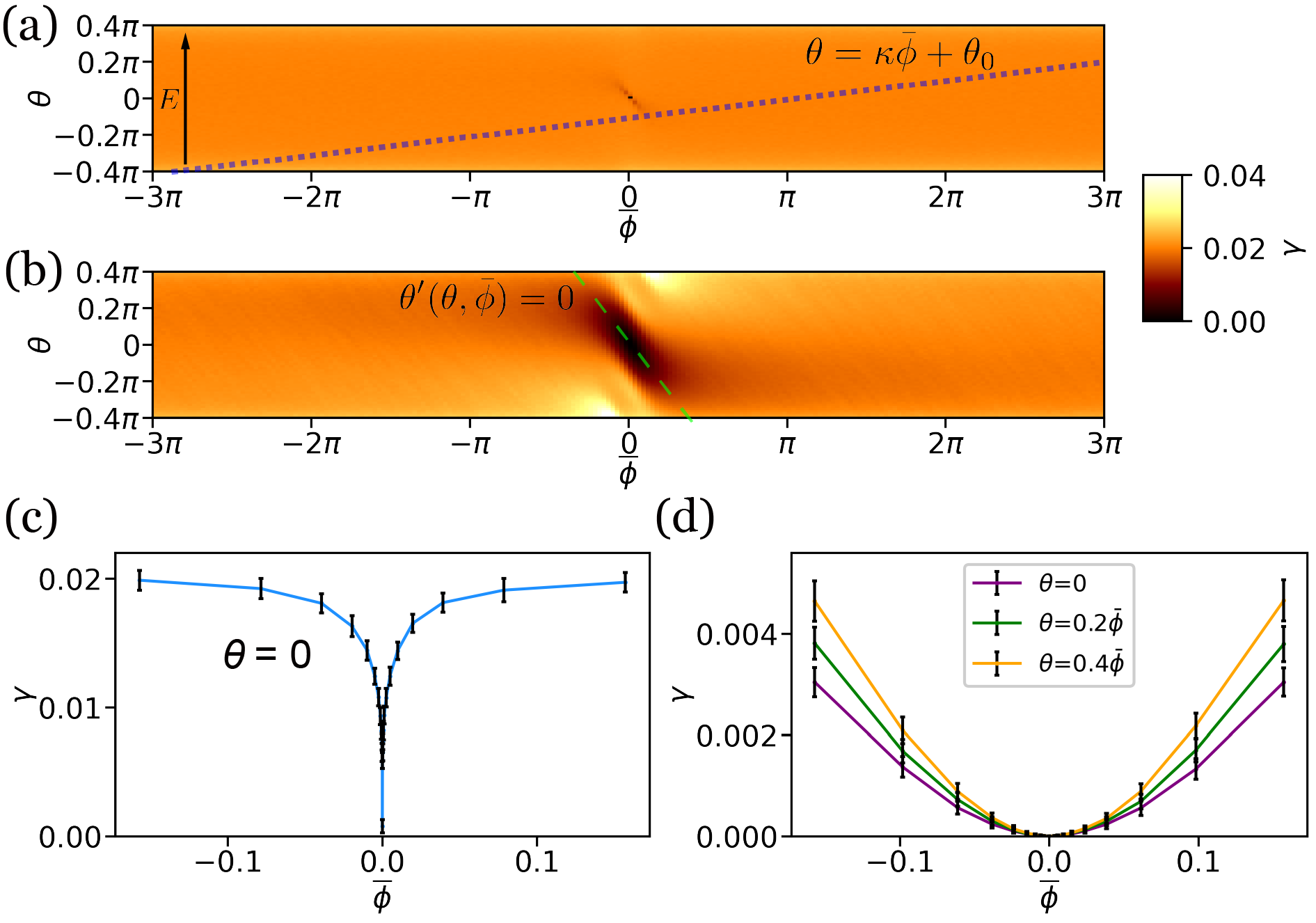}
    \caption{Computed Laypunov exponent for (a) ensemble I and  (b) ensemble II. $\gamma$ are averaged over $100$ samples with $2^{15}$ domain walls for each point in $\bar{\phi}-\theta$ plane, with $u_0=0.1e^{i\pi/8}$. Dashed lines indicate $\theta = \kappa\bar{\phi}+\theta_0$ (blue) and $\theta'=0$ (green). Critical behavior of $\gamma$ are shown for (c) ensemble I and (d) ensemble II. Vertical bars represent the standard deviations.} 
    \label{fig:05}
  \end{figure}

Ensemble I represents the non-interacting limit, where the domain wall magnetizations are uncorrelated so that $\Uparrow$ and $\Downarrow$ appear entirely by chance. As plotted in Fig.\ref{fig:05}(a), $\gamma$ in ensemble I is computed as a function of $\theta$ and $\bar \phi$, with $u=u_0/\cos\theta$ to account for  increasing back-scattering by a bond defect with larger $\theta$. Since the low-energy surface modes of the perfect FM bulk and defect zone superlattice show linear dispersion $E=v_0 k_z+\omega_0=v_d \theta+\omega_d$, $\bar{\phi}$ and $\theta$ are also related linearly, i.e. $\theta= \kappa \bar\phi+\theta_0$. $\kappa \bar{d}$ and $\theta_0$ are fixed by 
the model parameter specification which determines the values of Fermi velocities $v_0$, $v_d$ and Dirac points energy $\omega_0$, $\omega_d$.
Typically, $\gamma$ values are seen to be finite over the energy range examined, indicating a generic localization of the surface states of a domain wall-decorated AFM TI when the domain wall magnetizations are uncorrelated. 

The configurations in ensemble I generally correspond to $\mathcal{I}\neq \mathcal{I}^*$. 
Although $n_{\Uparrow}-n_{\Downarrow}=0$ over the entire sample, $|n_{\Uparrow}-n_{\Downarrow}|$ is of order $O(\sqrt{L})$ over a segment with $L$ domain walls due to statistical fluctuation.
This means that typically $|\alpha_i|\sim O(\sqrt{n})$, and the fluctuation in Ising moment $\mathcal{I}-\mathcal{I}^*\sim(\overline{\alpha^2}/n)^{1/2}\sim O(1)$, with  $\overline{\alpha^2}=\frac{1}{n}\sum_i\alpha^2_i$. Since the excess magnetic moments are carried by domain walls, $\alpha_i$ provides a measure of the macroscopic magnetization fluctuation. Consequently, $\mathcal I$ fails to converge in the thermodynamic limit, due to the unbounded magnetization fluctuation $|\alpha_i|$, signaling macroscopic breaking of time-reversal symmetry.

It follows immediately that if the fluctuation $|\alpha_i|$ is bounded in the thermodynamic limit, then $\mathcal{I}\rightarrow \mathcal{I}^*$ and the surface states are expected to be gapless. This might correspond to the physical situation for two considerations, namely,  the mediated AFM exchange interactions between domain walls that favors cancellation of magnetic moments, and the magnetic dipole interaction that suppresses magnetization on macroscopic scales\citep{Ashcroft76}. For a demonstration, domain wall sequences in ensemble II are generated using only three kinds of segments, "$\Uparrow\Downarrow$", "$\Uparrow\Uparrow\Downarrow\Downarrow$" ,
"$\Uparrow\Uparrow\Uparrow\Downarrow\Downarrow\Downarrow$", and their cyclic permutations. 
In this case $|\alpha_i|\le 3$ is bounded. Numerical results in Fig.\ref{fig:05}(b) show an oval-shaped region with vanishing $\gamma$ near $\theta=\overline{\phi}=0$ tilted along $\theta'=0$, which corresponds to the surface Dirac point of these $\mathcal{I}=\mathcal{I}^*$ configurations. The vanishing $\gamma$ (i.e. diverging localization length) suggests a delocalization transition
at the Dirac point in ensemble II when the model line $\theta= \kappa \bar\phi+\theta_0$ crosses the oval region.


\textcolor{blue}{\textit{Discussions.}} 
The computed delocalization transition near the Dirac point is closely related that in a 1D random hopping model due to an emergent chiral symmetry at half filling.\cite{balents1997,steiner1999,evers2008} A configuration in ensemble II is comprised of segments with zero magnetization, whose transfer matrices $\{M_i\}$ satisfy an effective time-reversal symmetry $\sigma_1 M_i \sigma_1=M_i^*,\; \forall i$.\cite{mello1991} Consequently, the random matrices $\{M_i\}$ conform to the orthogonal symmetry class\cite{evers2008} and equivalently describe the solution of a 1D stochastic Dirac equation\cite{suppl,comtet2010} 
\begin{equation}
    [-\mathrm i\sigma_3\partial_x +V_{E}(x)+m_{E}(x)\sigma_1]\Psi=0,
    \label{eq:Dirac}
\end{equation} 
which is invariant under $\sigma_1\mathcal K$ ($\mathcal K$ for complex conjugation). 
Here, the energy-dependent potential $V_E(x)$ and  mass $m_E(x)$ arise from stochastic pointer scatterers described by $M_i$. This model is known to possess delocalized solutions when $V_{E}(x)=0$ and $\langle m_{E}(x)\rangle=0$.\cite{balents1997,mathur1997,steiner1999,evers2008} Indeed, the oval region of delocalization along $\theta'=0$ for ensemble II  (FIG.\ref{fig:05}(b)) is close to the delocalization criticality of the Dirac equation: on the one hand, $\theta'=0$ puts the energy at the Dirac point in ensemble II, which corresponds to the Dirac point of Eq.(\ref{eq:Dirac}) (e.g. when $m_E=0$) appearing at $\langle V_{E}(x)\rangle=0$; on the other hand,  $\mathcal I-\mathcal I^*=0$ implies $\langle m_E(x)\rangle=0$ at the energy where $\theta'=0$, rendering the solution of Eq.(11) also gapless.

The mapping to Eq.(\ref{eq:Dirac}) is also feasible for ensemble I.~\cite{suppl} 
But since $\mathcal{I}-\mathcal{I}^*\neq 0$ generally indicates $\langle m_{E}(x)\rangle\neq 0$, ensemble I typically shows localization, except at $\theta=k_z=0$ when the surface spectra become chiral symmetric ($\omega_0=\omega_d$) and the total transfer matrix is an identity matrix. For the chiral symmetric models with $\theta_0=0$, the dip in the $\gamma$-$\bar \phi$ plot for ensemble I displays a standard critical scaling $\gamma \propto 1/\log|E-E_c|$ (see Fig.\ref{fig:05}(c) and note $\bar{\phi}\sim E$), as is also expected from a Eq.(\ref{eq:Dirac}) that describes a chiral symmetric Dirac Hamiltonian (i.e. when $V_E(x)=E_c-E$).\cite{balents1997,evers2008,ramola2014} A caveat is that at $E=E_c$, $\gamma$ may not be interpreted as the inverse localization length owing to large fluctuations,\cite{Texier10, ramola2014, balents1997,mathur1997,evers2008} as is also seen in our numerical results in FIG.\ref{fig:05}(c). In stark contrast, Fig.\ref{fig:05}(d) shows the delocalization transition in ensemble II exhibits an unusual algebraic scaling $\gamma \propto |E-E_c|^\nu$ ($\nu\sim 1.76$), and a substantial energy range with vanishing $\gamma$. Moreover, the sample-to-sample fluctuation in $\gamma$ is suppressed toward the critical point, pointing to an intriguing scenario that a sample adopting ensemble II turns out to be a ``good conductor'' at sufficiently low energies on its topological surfaces.

Experimentally, these results indicate a few intriguing phenomena to be studied 
on the surfaces of layered AFM TIs, concerning especially the $\mathrm{MnBi_{2n}Te_{3n+1}}$ family~\citep{zhang2015,otrokov2019,zhang2019b,deng2020a,ando2015,li2019}. We emphasize that our results apply generally to any topological surface preserving a $S$-symmetry and a $M\Theta$-symmetry with mirror $M$ parallel to $z$.\cite{suppl} Spectroscopic measurements may be employed to monitor the bandgap on any topological surfaces satisfying the above condition, and it is clearly interesting to find out whether such surfaces are gapped. Transport measurement will reveal the critical behavior of a material realization of the crossover from orthogonal to chiral orthogonal ensemble, where the unusual scaling and conductance fluctuation also offer valuable information on the magnetic correlation among AFM domain walls in the bulk material.


\begin{acknowledgments}
\textcolor{blue}{\textit{Acknowledgements.}} 
We are grateful for stimulating discussions with X.C. Xie and H. Jiang. We acknowledge the financial support from the National Natural Science Foundation of China (Grants No. 11725415 and 11934001), the National Key R\&D Program of China (Grants No.2018YFA0305601 and 2021YFA1400100), and the Strategic Priority Research Program of Chinese Academy of Sciences (Grant No. XDB28000000).
\end{acknowledgments}

\bibliographystyle{./statto}

\begin{thebibliography}{}

\end{thebibliography}


\begin{thebibliography}{10}

    \bibitem{zhang2015}
    Zhang, R.-X. \& Liu, C.-X.
    \newblock Topological magnetic crystalline insulators and corepresentation
      theory.
    \newblock \emph{Phys. Rev. B}, \textbf{91}, 115317 (2015).
    
    \bibitem{otrokov2019}
    Otrokov, M.~M. \emph{et~al.}
    \newblock Prediction and observation of an antiferromagnetic topological
      insulator.
    \newblock \emph{Nature}, \textbf{576}, 416--422 (2019).
    
    \bibitem{li2019}
    Li, J. \emph{et~al.}
    \newblock Intrinsic magnetic topological insulators in van der {{Waals}}
      layered {{MnBi}}{\textsubscript{2}}{{Te}}{\textsubscript{4}} -family
      materials.
    \newblock \emph{Sci. Adv.}, \textbf{5}, eaaw5685 (2019).
    
    \bibitem{zhang2019b}
    Zhang,D., Shi,M., Zhu,T., Xing,D., Zhang,H. \& Wang,J.
    \newblock Topological {{Axion States}} in the {{Magnetic Insulator
      MnBi}}$_2${{Te}}$_4$ with the {{Quantized Magnetoelectric Effect}}.
    \newblock \emph{Phys. Rev. Lett.}, \textbf{122}, 206401 (2019).
    
    \bibitem{liu2020a}
    Liu, C. \emph{et~al.}
    \newblock Robust axion insulator and {{Chern}} insulator phases in a
      two-dimensional antiferromagnetic topological insulator.
    \newblock \emph{Nature Mater.}, \textbf{19}, 522 (2020).
    
    \bibitem{deng2020a}
    Deng, Y. \emph{et~al.}
    \newblock Quantum anomalous {{Hall}} effect in intrinsic magnetic topological
      insulator {{MnBi}}$_2${{Te}}$_4$.
    \newblock \emph{Science}, \textbf{367}, 895 (2020).
    
    \bibitem{ge2020}
    Ge, J. \emph{et~al.}
    \newblock High-{{Chern-number}} and high-temperature quantum {{Hall}} effect
      without {{Landau}} levels.
    \newblock \emph{Natl. Sci. Rev.}, \textbf{7}, 1280 (2020).
    
    \bibitem{vidal2019a}
    Vidal, R.~C. \emph{et~al.}
    \newblock Surface states and {{Rashba-type}} spin polarization in
      antiferromagnetic {{MnBi}}$_2$ {{Te}}$_4$ (0001).
    \newblock \emph{Phys. Rev. B}, \textbf{100}, 121104(R) (2019).
    
    \bibitem{hao2019}
    Hao,Y.J., Liu,P., Feng,Y., Ma,X.M., Schwier,E.F., Arita,M. \emph{et~al.}
    \newblock Gapless {{Surface Dirac Cone}} in {{Antiferromagnetic Topological
      Insulator MnBi}}$_2$ {{Te}}$_4$.
    \newblock \emph{Phys. Rev. X}, \textbf{9}, 041038 (2019).
    
    \bibitem{swatek2020a}
    Swatek,P., Wu,Y., Wang,L.L., Lee,K., Schrunk,B., Yan,J.\& Kaminski,A.
    \newblock Gapless {{Dirac}} surface states in the antiferromagnetic topological
      insulator {{MnBi}}$_2$ {{Te}}$_4$.
    \newblock \emph{Phys. Rev. B}, \textbf{101}, 161109(R) (2020).
    
    \bibitem{ando2015}
    Ando, Y. \& Fu, L.
    \newblock Topological crystalline insulators and topological superconductors:
      {{From}} concepts to materials.
    \newblock \emph{Ann. Rev. Cond. Matt. Phys.}, \textbf{6}, 361 (2015).
    
    \bibitem{zhang2009}
    Zhang, H. \emph{et~al.}
    \newblock Topological insulators in {{Bi$_2$Se$_3$}}, {{Bi$_2$Te$_3$}} and
      {{Sb$_2$Te$_3$}} with a single {{Dirac}} cone on the surface.
    \newblock \emph{Nat. Phys.}, \textbf{5}, 438 (2009).
    
    \bibitem{zhang2020}
    Zhang, R.-X., Wu, F. \& Das~Sarma, S.
    \newblock M\"obius {{Insulator}} and {{Higher}}-{{Order Topology}} in
      {{MnBi$_{2n}$Te$_{3n+1}$}}.
    \newblock \emph{Phys. Rev. Lett.}, \textbf{124}, 136407 (2020).
    
    \bibitem{suppl}
    Supplemental materials.
    
    \bibitem{sancho1985}
    Sancho, M. P.~L., Sancho, J. M.~L., Sancho, J. M.~L. \& Rubio, J.
    \newblock Highly convergent schemes for the calculation of bulk and surface
      {{Green}} functions.
    \newblock \emph{J. Phys., F Met. Phys.}, \textbf{15}, 851 (1985).
    
    \bibitem{Goldsheid89}
    Gol{\textquotesingle}dsheid, I.~Y. \& Margulis, G.~A.
    \newblock \href {https://doi.org/10.1070/rm1989v044n05abeh002214} {Lyapunov
      indices of a product of random matrices}.
    \newblock \emph{Rus. Math. Surv.}, \textbf{44}, 11 (1989).
    
    \bibitem{Kramer93}
    Kramer, B. \& MacKinnon, A.
    \newblock Localization: theory and experiment.
    \newblock \emph{Rep. Prog. Phys.}, \textbf{56}, 1469 (1993).
    
    \bibitem{Pichard86}
    Pichard, J.~L. \& Andr{\'{e}}, G.
    \newblock \href {https://doi.org/10.1209/0295-5075/2/6/011} {Many-channel
      transmission: Large volume limit of the distribution of localization lengths
      and one-parameter scaling}.
    \newblock \emph{Europhys. Lett.}, \textbf{2}, 477 (1986).
    
    \bibitem{Geist90}
    Geist, K., Parlitz, U. \& Lauterborn, W.
    \newblock Comparison of {{Different Methods}} for {{Computing Lyapunov
      Exponents}}.
    \newblock \emph{Prog. Theo. Phys.}, \textbf{83}, 875 (1990).
    
    \bibitem{Ashcroft76}
    Ashcroft, N. \& Mermin, N.
    \newblock \emph{{Solid State Physics}}.
    \newblock Saunders College, Philadelphia (1976).
    
    \bibitem{balents1997}
    Balents, L. \& Fisher, M. P.~A.
    \newblock Delocalization transition via supersymmetry in one dimension.
    \newblock \emph{Phys. Rev. B}, \textbf{56}, 12970 (1997).
    
    \bibitem{steiner1999}
    Steiner, M., Chen, Y., Fabrizio, M. \& Gogolin, A.~O.
    \newblock Statistical properties of a localization-delocalization transition in
      one dimension.
    \newblock \emph{Phys. Rev. B}, \textbf{59}, 14848 (1999).
    
    \bibitem{evers2008}
    Evers, F. \& Mirlin, A.~D.
    \newblock Anderson transitions.
    \newblock \emph{Rev. Mod. Phys.}, \textbf{80}, 1355 (2008).
    
    \bibitem{mello1991}
    Mello, P.~A. \& Pichard, J.-L.
    \newblock Symmetries and parametrization of the transfer matrix in electronic
      quantum transport theory.
    \newblock \emph{J. Phys. I}, \textbf{1}, 493 (1991).
    
    \bibitem{comtet2010}
    Comtet, A., Texier, C. \& Tourigny, Y.
    \newblock Products of {{Random Matrices}} and {{Generalised Quantum Point
      Scatterers}}.
    \newblock \emph{J. Stat. Phys.}, \textbf{140}, 427 (2010).
    
    \bibitem{mathur1997}
    Mathur, H.
    \newblock Feynman's propagator applied to network models of localization.
    \newblock \emph{Phys. Rev. B}, \textbf{56}, 15794 (1997).
    
    \bibitem{ramola2014}
    Ramola, K. \& Texier, C.
    \newblock Fluctuations of random matrix products and {{1D Dirac equation}} with
      {{random mass}}.
    \newblock \emph{J. Stat. Phys.}, \textbf{157}, 497 (2014).
    
    \bibitem{Texier10}
    Texier, C. \& Hagendorf, C.
    \newblock \href {https://doi.org/10.1088/1751-8113/43/2/025002} {The effect of
      boundaries on the spectrum of a one-dimensional random mass {Dirac
      Hamiltonian}}.
    \newblock \emph{J. Phys. A Math. Theor.}, \textbf{43}, 2, 025002 (2009).
    
    \end{thebibliography}

\end{document}